# Exotic behavior and crystal structures of calcium under pressure

**Artem R. Oganov[1,2], Ying Xu[3], Yanming Ma[3], Ion Errea[4,5], Aitor Bergara[4,5,6] and Andriy O. Lyakhov[1]**

[1]Department of Geosciences, Department of Physics and Astronomy, and New York Center for Computational Sciences, Stony Brook University, Stony Brook, NY 11794-2100, USA,
[2]Geology Department, Moscow State University, 119992 Moscow, Russia,
[3]National Lab of Superhard Materials, Jilin University, Changchun 130012, P.R. China,
[4]Materia Kondentsatuaren Fisika Saila, Zientzia eta Teknologia Fakultatea, Euskal Herriko Unibertsitatea, 644 Postakutxatila, 48080 Bilbo, Basque Country, Spain,
[5]Donostia International Physics Center (DIPC), Paseo de Manuel Lardizabal, 20018, Donostia, Basque Country, Spain,
[6]Centro Fisica de Materiale CSIC-UPV/EHU, 1072 Posta kutxatila, E-20080 Donostia, Basque Country, Spain.

**Experimental studies established that calcium undergoes several counterintuitive transitions under pressure: fcc → bcc → simple cubic → Ca-IV → Ca-V, and becomes a good superconductor in the simple cubic and higher-pressure phases. Here, using** *ab initio* **evolutionary simulations, we explore the behavior of Ca under pressure and find a number of new phases. Our structural sequence differs from the traditional picture for Ca, but is similar to that for Sr. The β-tin (***$I4_1/amd$***) structure, rather than simple cubic, is predicted to be the theoretical ground state at 0 K and 33-71 GPa. This structure can be represented as a large distortion of the simple cubic structure, just as the higher-pressure phases stable between 71 and 134 GPa. The structure of Ca-V, stable above 134 GPa, is a complex host-guest structure. According to our calculations, the predicted phases are superconductors with** *Tc* **increasing under pressure and reaching ~20 K at 120 GPa, in good agreement with experiment.**

\body

## Introduction

Calcium exhibits a non-trivial and somewhat mysterious behavior under pressure. At 19.5 GPa it transforms from the face-centred cubic (fcc) to the body-centred cubic (bcc) structure, and then, at 32 GPa, to the simple cubic (sc) structure [0,2]. Such a sequence of transitions is exactly opposite to normal intuition, as it is accompanied by a decrease of coordination numbers (12 → 8 → 6) and sphere packing efficiency (0.74 → 0.68 → 0.52). Good metal at ambient conditions, fcc-Ca shows increasing electrical resistivity under pressure [3-5]. Even more intriguingly, the resistivity of the fcc phase at and just below this maximum has negative temperature derivative, characteristic of the non-metallic (semiconducting) state, consistent with a small band gap found in *ab initio* calculations [6] in the same pressure range. Such *demetallization* under pressure is counterintuitive: since at sufficiently high pressure all materials must become free-electron metals, the expected behavior is an increasing tendency to the free-electron limit under pressure (see [7] for a discussion). Contrary to these expectations, strong departure from the free-electron state under pressure has also been found for sodium [8,9] and lithium [10-12] at megabar pressures.

*Ab initio* calculations [13] confirmed the fcc → bcc → sc structure sequence and yielded reasonably accurate values for the transition pressures. However, the sc phase encounters problems: it cannot be explained within the Hume-Rothery approach (Fermi surface – Brillouin zone interaction) unless one assumes 4 valence electrons per atom [14], and, even more seriously, lattice dynamics calculations [15,16] showed that it is dynamically unstable, and although this dynamical instability may be lifted by anharmonic effects [16,17], other structures (see below) have much lower enthalpies. This seeming contradiction with experiments that initially showed a perfect sc structure [0,2,18] was largely resolved by recent experiments [17] showing that the sc structure in reality is indeed distorted and that the observed low-temperature behaviour of calcium is likely affected by metastability.

Remarkably, Ca is a superconductor above 50 GPa [19], and its *Tc* rapidly (and with an increasing slope) soars with pressure, reaching 25 K at 161 GPa [20] - the highest *Tc* value found in any element. While already the sc phase is superconducting, the most intriguing high *Tc* values are found in the stability fields of other phases; a recent X-ray diffraction study [2] found two further phases beyond sc - Ca-IV (stable at 113-139 GPa) and Ca-V (stable above 139 GPa). First structural models, proposed in [21], gave a structure possessing the $P4_32_12$ space group and 8 atoms in the unit cell (here denoted as $P4_32_12$-8) for Ca-IV and *Cmca*-8 for Ca-V. The $P4_32_12$-8 structure, proposed for Ca-IV, has indeed been confirmed experimentally [22]. A very recent theoretical study made a different proposition - a *Pnma*-4 structure for Ca-IV and the same *Cmca*-8 structure for Ca-V [23]. However, using evolutionary global optimization techniques, for Ca-V we found (see below) a self-hosting structure, more stable than the *Cmca*-8 structure proposed in [21], and providing an excellent match to experimental diffraction data and observed stability field of Ca-V.

Here we predict and examine new crystal structures and address the counterintuitive behavior of calcium under pressure - the apparent decrease of its packing efficiency, transition into the semiconducting state at moderate pressures, link between structure and superconductivity, stability of self-hosting structures under pressure, and unusually high *Tc* values observed for calcium. All these aspects make calcium one of the most anomalous and interesting elements in the Periodic Table.

## Few remarks on "simplicity" of calcium

Before discussing the new structures predicted here, let us briefly consider two aspects of non-trivial behavior of calcium at moderate pressures (< 50-100 GPa) - (i) the apparent opening of the band gap below the fcc → bcc transition (19.5 GPa), and (ii) the anomalous fcc → bcc → sc transition sequence.

With an even number (two) of valence electrons in the primitive cell, calcium could be an insulator - it is a metal only due to band overlap (or, in real-space language, due to the spatial extent of the wavefunctions being much larger than the shortest interatomic distance). Very compressible, the atomic volume of calcium shrinks by a factor of 2.4 on going from the atmospheric pressure to 50 GPa (see theoretical and experimental equations of state in [6]), inducing large changes in the electronic structure. While in the isolated calcium atom only s- and p-orbitals are occupied (it is the last element in the Periodic Table without d-electrons), d-orbitals are low-energy and can be populated under pressure - indeed, theory predicts an s → d transition in calcium and all heavy alkali earth and alkali metals [13,6,24]. Compression has a different effect on the energies of different orbitals (or bands), and generally higher-$l$ orbitals that are unoccupied in the free atom become populated under pressure. In calcium's group I neighbour, potassium, d-electrons become dominant at the Fermi level already at ~10 GPa (e.g., ref. 24), i.e. slightly earlier than in calcium. The s→d transition greatly affects reactivity of potassium and may enable it to alloy with Fe in the Earth's core [25] (this is geochemically very important, as $^{40}$K is an important source of radiogenic heat in the Earth and the only possible radioactive element in the core); similarly large changes in reactivity may also occur in calcium.

Given that the ionic radius of $Ca^{2+}$ is 1.0 Å (which is an effective core radius) and the 4s-orbital radius is 1.69 Å, from the equation of state of calcium we deduce that the (interatomic) core-valence overlap is large at all pressures above 30 GPa, and core-core overlap becomes significant at 300 GPa. Overlaps of valence and core orbitals of one atom with core orbitals of another atom can lead to extremely interesting physical effects, such as expulsion of valence electrons into the "empty" space of the structure leading to the formation of strong non-nuclear charge maxima and possible demetallization [12,26,8].

The degree of localization of these electron pairs increases with pressure, which explains the observed increase of the resistivity and semiconducting behavior of fcc calcium under pressure. This is seen also in a very non-free-electron-like electronic density of states (Fig. 1), showing a small band gap (~ 0.1 eV) at 18 GPa. The increase of the non-nuclear density maxima under pressure is correlated with a rapid increase of d-orbital occupancy $n_d$: projecting wavefunctions inside atom-centred spheres of 2.0 Å radius, we find $n_d$ = 0.36 at 1 atm and $n_d$ = 0.83 at 18 GPa (see also [27,13,6]). Across all alkali earth metals $n_d$ > 0.92 appears to be a good phenomenological criterion for the onset of superconductivity [27].

Fig. 1 shows an unusual distribution of the valence electron localization function (ELF - [28]) in fcc-Ca at 1 atm: it has maxima not only at the nuclei, but also in the octahedral voids between them (thus, ELF maxima form a NaCl-type structure). These maxima become much more pronounced on increasing pressure, and their origin can be traced to the exclusionary effect of the core electrons on the valence electrons, first predicted to occur in lithium [12,26] and sodium [9,8].

The fcc → bcc → sc structural sequence is less anomalous than often thought. The sphere packing efficiencies (i.e. ratio of the volume occupied by touching atomic spheres to the total volume) of the fcc, bcc and sc structures are 74%, 68% and 52%, respectively. If atoms were of the same size (or bonds of the same length) in all structures, fcc would be the densest. However, in reality the bcc and sc structures are denser (at transition pressures) than fcc because of the shorter interatomic distances. Greater density of the bcc and sc structures has often been explained by the s → d electronic transition; this is not necessary and even incorrect: indeed, magnesium (not undergoing any electronic transitions) has a similar transition (hcp → bcc) at 50 GPa [29]. Generally, lower-coordination structures can be denser than fcc or hcp even in the absence of electronic transitions – because lower coordination corresponds to shorter bonds. Indeed, atomic sizes and bond lengths $R$ depend on the coordination number ν [30]:

$$R = R_0 + b \ln \nu \tag{1}$$

where $R$ and $R_0$ are expressed in Å, $R_0$ is a bond-specific constant and $b$ = 0.37 Å. Using (1), the ratio of atomic volumes in the fcc and bcc structures at 1 atm is:

$$\frac{V_{bcc}}{V_{fcc}} = \frac{f_{bcc}}{f_{fcc}} \left( \frac{R_{fcc} + b \ln(8/12)}{R_{fcc}} \right)^3 \tag{2}$$

where $f$ are the packing efficiencies, and $R_{fcc}$ is the bond length in the fcc structure. This expression does indeed show that in very many cases, without any need for electronic transitions, the bcc structure can be denser than fcc or hcp at 1 atm, and more stable under pressure. While the hard-sphere model fails completely (Fig. 2), model (2) is more consistent with the computed atomic volumes, but still does not give quantitative agreement (to achieve which one might need a more complicated model including higher coordination spheres and delocalized electrons). Fig. 2 shows that in most cases bcc and fcc structures have similar densities, with bcc being slightly denser. Thus, the fcc → bcc transition is not an anomaly. On the other hand, the sc structure is usually much less dense and only for a handful of elements (Li, C, Ba) is denser than fcc or bcc. Note that for calcium at 1 atm the sc structure is *slightly less dense* than fcc (Fig. 2) and becomes denser under pressure due to its higher compressibility.

**Results of evolutionary simulations: new phases of calcium**

Evolutionary simulations at 20 GPa and 30 GPa found the bcc structure to be stable, in agreement with experiment [0,2,18]. The lowest-enthalpy structure found at 40 GPa and 70 GPa is, however, not sc, but an *I*4$_1$/*amd* structure (β-tin type), which has recently been reported in theoretical studies [31,32] and can be described as a strongly distorted sc structure. Fig. 3 depicts how the β-tin structure is favored against sc and becomes the theoretical ground state above 33 GPa, exactly at the pressure at which very recent experiments found that fcc nor bcc phase are no longer stable [18]. Previous theoretical works [15,16,33] that found sc structure to be dynamically unstable associated to strong nestings in its Fermi surface. At 40 GPa the $I4_1/amd$ phase is slightly (0.9%) less dense than the sc-structure, but significantly (52 meV/atom) more favorable.

This energy difference between β-tin and sc is not expected to be overcome by the zero point energy (ZPE) considering that in our calculations at 50 GPa the ZPE in the latter phase is just 5 meV/atom lower. As shown in Fig. 4, the coordination of Ca atoms is octahedral (as in the sc phase), but distorted - with 4 nearest neighbors at 2.68 Å, and 2 neighbors at 2.77 Å, suggesting a Jahn-Teller or Peierls distortion (at this pressure, Ca atoms have an electronic configuration close to $s^{0.5} p^{0.5} d^1$, which is expected to show the Jahn-Teller effect). This distortion opens a pseudogap which decreases the density of states at the Fermi level and lowers the electronic kinetic energy of the preferred $I4_1/amd$ phase over the sc. The same structure type is known for other elements - Sn, and high-pressure forms of Ge, Si. Most interestingly, it is adopted by the phase III of strontium in the pressure range 26-35 GPa [34,35].

At 71 GPa the β-tin structure distorts into a $C2/c$-12 (Sr-IV) structure with 12 atoms in the conventional unit cell, and this structure remains stable up to 89 GPa. C2/c-12 (Sr-IV) structure is a helical distortion of the β-tin structure with a triplication of the unit cell size. The same helical distortion of the β-tin structure was reported in Sr-IV phase (ref. 35, where the space group, however, was misdetermined as $Ia$ – while a closer inspection shows it to be $C2/c$). Up to this point, the observed [34] sequence of phase transitions in strontium and the predicted sequence in calcium are identical: fcc → bcc → $I4_1/amd$ → $C2/c$-12 (Sr-IV). The next higher-pressure phase of strontium, Sr-V, is an incommensurate host-guest structure. We do find such a structure also for calcium, but preceded by two other phases. On increasing pressure to ~ 100 GPa, we find extremely distorted versions of the simple cubic structure. One of these is a metastable (but very competitive) $Cmca$-16 structure, which can be described as a frustrated structure intermediate between sc and hexagonal close-packed structures. Fig. 3 shows that in the pressure range 100-130 GPa there are several energetically extremely close structures, and the most stable structure at 89-116 GPa is the $P4_32_12$-8 structure proposed by Ishikawa et al. [25], this structure also can be derived from sc by a large distortion and has recently been experimentally confirmed for Ca-IV phase [26]. The $Pnma$-4 structure is the stable phase at higher pressures, 116-134 GPa, and is the last stable structure related to sc. The structure proposed in [25] for Ca-V (also belonging to the family of sc-derived structures) is never stable at $T = 0$ K.

The structure we found for Ca-V is a host-guest structure, similar to Sr-V and Ba-IV [36] (Fig. 5). This result is similar to that recently reported by Arapan et al. [37], who found this structure by educated guess based on a possible analogy with Sr. Such phases for Sr and Ba are incommensurate - since incommensurate structures cannot be predicted within strictly imposed periodic boundary conditions, we can only produce their commensurate approximants. In fact, our evolutionary simulations yielded several energetically nearly degenerate and geometrically very similar structures - such as $C2/m$-32 (host-guest) (the most stable structure) and $I4/mcm$-32 (host-guest) and $C2/c$-32 (host-guest) structures. The reader is referred to the supporting online information for specific details on these structures. This near degeneracy suggests that the system is frustrated by competing interactions – a common reason behind incommensurate phases. In experiments [2], Ca-V was seen coexisting with Ca-IV above 139 GPa - this metastable coexistence implies a large energy barrier for the phase transition, likely due to large structural differences between Ca-IV and Ca-V. This is consistent with our results, whereas in [21] both proposed structures are rather similar derivatives of the sc structure. We also note that since Ca-IV is related to the sc structure, finding it with a neighborhood search method (metadynamics, starting from the sc structure, was used in [21]) is very efficient, while Ca-V would be more challenging to neighborhood algorithms and a fully global search based on evolutionary algorithms should be preferred.

Interestingly, all distortions of the simple cubic structure (β-tin, $C2/c$-12 (Sr-IV), $P4_32_12$-8 and $Pnma$-4), predicted to be stable below 134 GPa, have a highly coupled soft mode associated to the distortion as a major contributor to the observed superconductivity (see supporting information). For instance, the maximum phonon linewidth for the β-tin structure at 60 GPa corresponds to a strongly softened optical mode around the Γ-point. It is worth noting that such mode corresponds to the unstable anharmonic mode at M symmetry point [21] in the sc phase. Due to a Peierls distortion this mode is stabilized in the β-tin phase and becomes the one with largest partial contribution to the electron-phonon coupling constant λ. Additionally, Fig. 6 displays our calculated superconducting transition temperatures for these structures at 60, 80, 110, 120 and 130 GPa. The calculations closely agree with the experimental data [20] and reflect the pressure induced enhancement of $T_c$, reaching 23.5 K at 130 GPa, supporting the validity of the proposed structures.

**Discussion**

We have predicted a number of new phases of calcium, one of the most interesting and enigmatic elements in the Periodic System - an element that displays nearly all of the puzzling phenomena recently discovered in the elements (high-pressure superconductivity, host-guest structures, unexpected transitions accompanied by electronic transitions and a decrease of coordination numbers and packing efficiencies). The sequence of structures found here differs significantly from the traditional picture [34], but is very similar to the one known for strontium, the closest analogue of calcium. The main difference is that instead of a single sc phase we find a family of sc-derived structures, which are energetically more favorable. It was recently proposed [16,31,17] that the dynamical instability of the sc structure can be overcome by anharmonic effects even at quite low temperatures. This does not automatically imply thermodynamic stability and we suggest a reconsideration of the experimental evidence for the sc structure and search for additional phases predicted here at $T = 0$ K. The first step in this direction has been made in recent experiments [17].

The similarity of the several structures stable at 33-134 GPa (β-tin, $C2/c$-12 (Sr-IV), $P4_32_12$-8, $Pnma$-4) with the sc structure is evident not only from their appearance, but also from their X-ray diffraction profiles (Fig. 7) that bear great similarity to the diffraction pattern of the sc structure - but have split peaks due to lower symmetries. Since some extra peaks can become invisible due to possible texturing of the samples used in experiments, the β-tin structure could explain the experimental diffraction patterns ascribed to the simple cubic structure. However, recent single crystal X-ray diffraction study [17] favors slightly distorted versions of the sc structure instead, but suggests that these are metastable, the true ground state likely being the β-tin structure. We also note that our host-guest structure provides a better explanation of experimental data for the Ca-V phase than the model proposed in [21]. Our model has a single strong peak that corresponds to the extra peak found in [2], all other peaks are explained by Ca-IV. As discussed above, fcc-Ca possesses a number of chemical and physical anomalies related to the strong electron localization in the interstitial space. In all the higher-pressure phases the degree of interstitial electron localization progressively decreases, and can be considered negligible for all phases beyond the β-tin phase (pressures >71 GPa). We showed that the fcc → bcc transition is a normal phenomenon in the elements under pressure, due to the usually higher density of the bcc structure. Further transitions to the β-tin (or sc) derivative structures are more difficult to explain in simple terms. We note, however, that (i) several elements adopt such structures under pressure, (ii) the pressure of the bcc → sc-family transition corresponds to a large occupancy of the 3d-orbitals and a large core-valence interaction on neighboring atoms, and (iii) that the β-tin structure can be stabilised by the Jahn-Teller distortion enabled by the s→d electronic transition in calcium. The large number of these derivative structures and their frustrated topologies and similar enthalpies suggest the presence of competing factors – namely, bonding (which favors simple and symmetric structures) and density (the increase of which necessitates electronic transitions). Finally, at 134 GPa, an entirely new type of structures takes over - host-guest structures unrelated to the sc structure and displaying another type of geometric frustration related to the incommensurate modulation. The precise origins of such host-guest structures are still unclear: orbital characters of both host and guest are very similar and the material appears to be a good metal without any strong non-nuclear density maxima. The extent of the stability of the host-guest structure is not known yet - but both from present theory and experiment [20] it appears to be very large (from 134 GPa to at least 161 GPa). Our simulations show that the hcp structure (which succeeds the host-guest one for Ba [34] on increasing pressure) becomes stable only above 564 GPa. At such pressures the s→d transition is completed (as in the very similar double-hcp structures of K, Rb, Cs [26], which are formed at higher pressures than the complex incommensurate host-guest structures in these elements) and core-core overlap (not just core-valence) is significant. Complex behaviour of calcium at 32-564 GPa is a consequence of (i) competition of bonding (energetic) and density factors, (ii) gradual s→d transition (related to compression and core-valence overlaps) and two effects related to it – the Jahn-Teller distortion and s-d hybridization that gives rise to directional bonding. At pressures of stability of hcp-Ca (>564 GPa), the $PV$ term (>23.4 eV) far exceeds bond energies, and electronic kinetic energy is significantly greater than potential energy (this favors electronic delocalization and destruction of directional bonding and leads to a simple and symmetric hcp structure).

**Methods**
All calculations presented here are based on density functional theory within the generalised gradient approximation (GGA [38]) and using the all-electron PAW method [39]. Structure prediction calculations were done using an evolutionary algorithm developed by Oganov and Glass and implemented in the USPEX code [40-42], while lattice dynamics and superconducting properties were studied using density-functional perturbation theory (DFPT) [43] using the Quantum-ESPRESSO package [44].
Detailed evolutionary simulations were performed at 70 GPa, 100 GPa, 130 GPa and 160 GPa with system sizes of 3, 4, 6, 8, 9, 12 and 16 atoms in the unit cell. Exploratory runs with 8 atoms/cell were also done at 20 GPa, 30 GPa and 40 GPa, and with 16 atoms/cell at 300 GPa and 600 GPa. Each generation contained between 10 and 40 structures (increasing with the system size), and the first generation was always produced randomly. All structures were locally optimized using the VASP code [45]. In a typical evolutionary run, the lowest-enthalpy 60% of each generation were used to produce the next generation (70-80% of these were produced by heredity, and the remaining 20-30% by lattice mutation, and in addition to that the lowest-enthalpy structure survived into the next generation). Gaussian strength of lattice mutation [41] was set to 0.5-0.55. Local optimizations performed during structure search, were done with the conjugate gradients method and were stopped when the enthalpy changes became smaller than 1 meV/cell. The PAW potential used in these calculations has the [Ne] core (radius 2.3 a.u.). We used the plane wave cutoff of 350 eV, which proved to give excellent convergence of the stress tensor and energy differences. The Brillouin zone was sampled with Monkhorst-Pack meshes [46] of $2\pi \times 0.09$ Å$^{-1}$ resolution, and the Methfessel-Paxton electronic smearing scheme [47] with the electronic temperature of 0.1 eV. These settings proved to give excellent convergence for the stress tensor and energy differences between structures. Analyzing the results of evolutionary simulations, we selected a number of distinct lowest-enthalpy structures and optimized their structures as a function of pressure using VASP and very dense ($2\pi \times 0.03$ Å$^{-1}$ resolution) k-points meshes.
Electron-phonon coupling calculations were performed using DFPT, with an ultrasoft pseudopotential [48] treating $3s^23p^64s^2$ as valence electrons. Convergence tests showed that Monkhorst-Park k-point meshes of 16×16×16, 12×6×6, 9×9×3 and 8×8×8 for the $I4_1/amd$, $C2/c$-12 (Sr-IV), $P4_32_12$-8 and $Pnma$-4 structures respectively, were adequate for Brillouin zone integrations. The transition temperatures to the superconducting states were calculated making use of the Allen-Dynes modified McMillan equation [49] where the typical value of $\mu^* = 0.1$ was taken for the Coulomb pseudopotential.

**Acknowledgments**
A. R. O. gratefully acknowledges funding from the Research Foundation of Stony Brook University, Intel Corp., and Rosnauka (Russia, contract 02.740.11.5102), and access to the Skif MSU supercomputer (Moscow State University) and to the Joint Supercomputer Centre of the Russian Academy of Sciences. A. B. gratefully acknowledges financial support from the EC 6th framework Network of Excellence NANOQUANTA (NMP4-CT-2004-500198) and I. E. would like to thank the Basque Department of Education and Research for financial help and SGI-IZO SGIker UPV-EHU for the allocation of computational resources.

## FIGURE LEGENDS

**Fig. 1.** Valence ELF (including semicore 3s and 3p states) and density of states of fcc-Ca at (a,c) 1 atm and 18 GPa (b,d). ELF plots (a,b): (100) sections through positions of Ca atoms and non-nuclear electron density maxima ("$\bar{e}$"). Minimum/maximum ELF values are 0.05/0.73 in (a) and 0.08/0.84 in (b), with contour spacing of 0.05. Note ELF peaks in the core regions and in the interstices (valence electrons expelled from the core regions); at 18 GPa interstitial valence electrons are more localized than semicore electrons. In DOS plots, energies are relative to the Fermi level; note the narrow band gap at 18 GPa.

**Fig. 2.** Atomic volumes in the bcc (gray squares) and sc (black circles) phases, relative to those in the fcc structure. In the horizontal axis the elements are arranged by the metallic radius in the fcc phase. All results are based on present GGA calculations. While bcc phases have similar densities to fcc (and for most elements are even slightly denser), hypothetical sc phases are usually less dense - except C, Li and Ba. It is clear that predictions of the hard-sphere model are inconsistent with *ab initio* calculations.

**Fig. 3.** (a) Enthalpies of the bcc, sc and $I4_1/amd$ (β-tin phase) structures (relative to fcc). (b) Enthalpies of several competitive phases (relative to the β-tin phase). A few other structures ($C2/c$-32 (host-guest), $C2/c$-24, $I4/mcm$-32 (host-guest)), very nearly degenerate with the ground states, are not shown for clarity.

**Fig. 4.** Structures of a) simple cubic, b) β-tin-type tetragonal distortion ($I4_1/amd$) of the simple cubic structure for Ca at 40 GPa. The $I4_1/amd$ structure is significantly more favorable. Bond lengths are indicated.

**Fig. 5.** $C2/m$-32 (host-guest) structure at 150 GPa. 10-coordinate polyhedra (bicapped square antiprisms) are shown: a) view along the pseudotetragonal axis, b) side view showing packing of the polyhedra (it can also be seen that the genuine symmetry of this periodic approximant is lower than tetragonal). This structure can be described as self-hosting, the host structure has Ca in 8-fold coordination and makes a framework in which guest atoms occupy 10-coordinate sites. Other structures are illustrated in the supporting online material.

**Fig. 6.** Comparison of the calculated $T_c$ with the experimental values obtained by Yabuuchi et al. [20]. The calculated $T_c$ values show good agreement with experimental data across the whole pressure range.

**Fig. 7.** Summary of X-ray diffraction patterns. $\lambda = 0.6198$ Å. Structures $I4/mcm$-32 (host-guest) and $C2/c$-32 (host-guest) have very similar diffraction patterns to $C2/m$-32 (host-guest) shown here. The absence of several reflections, predicted for the β-tin structure at 40 GPa, in the experimental diffraction pattern, might be explained by preferred orientation with a unique [101] or [111] direction. The experimental diffraction pattern at 139 GPa is explained (as proposed in [2]) by a mixture of phases IV and V.

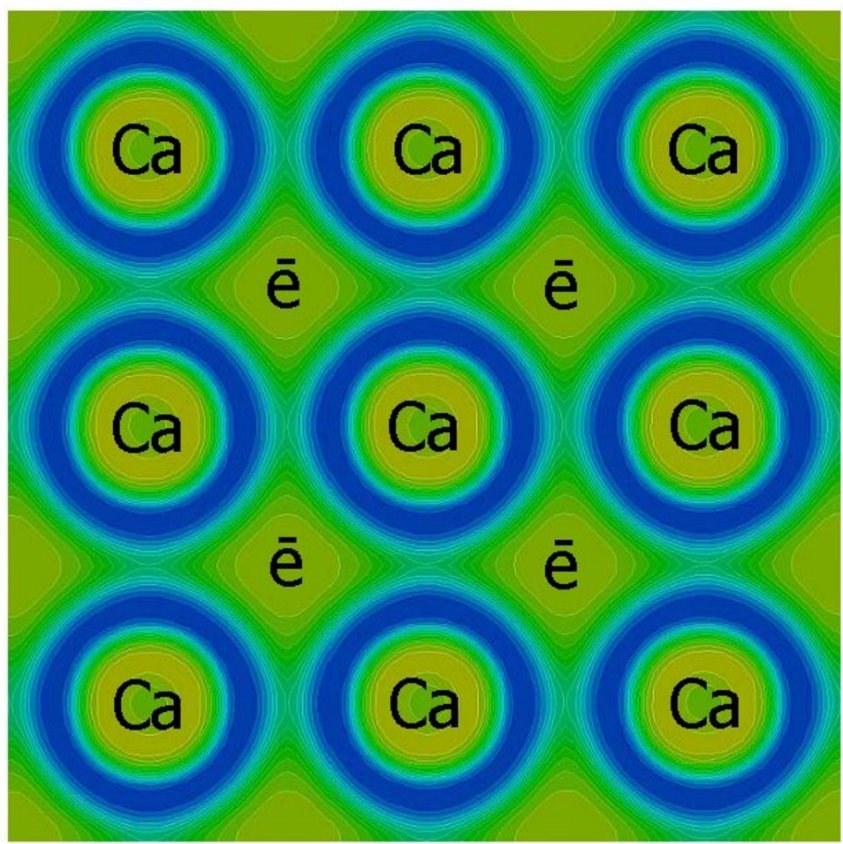 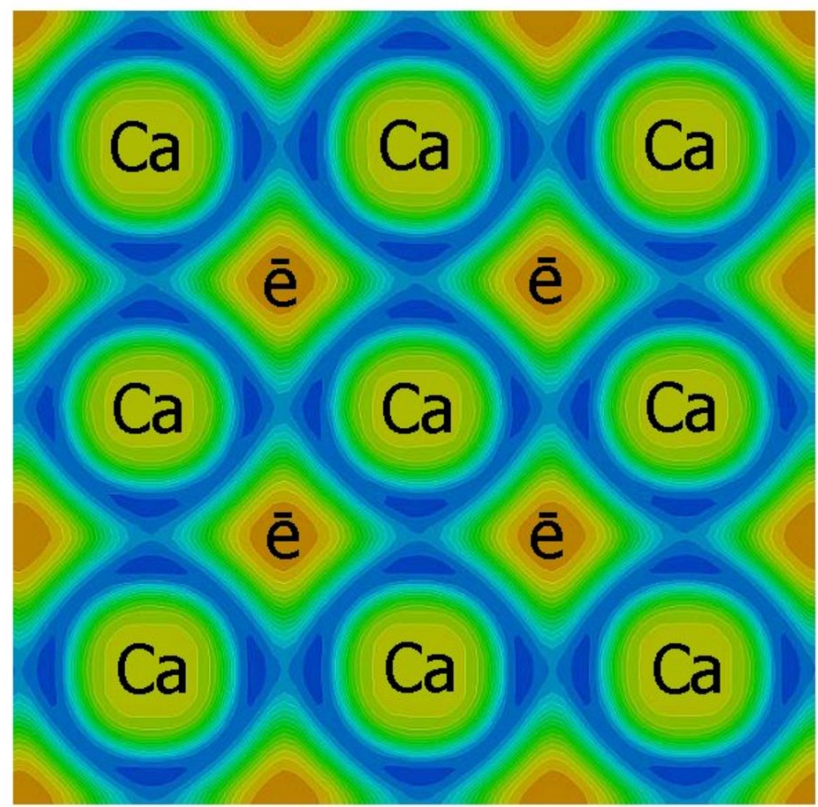
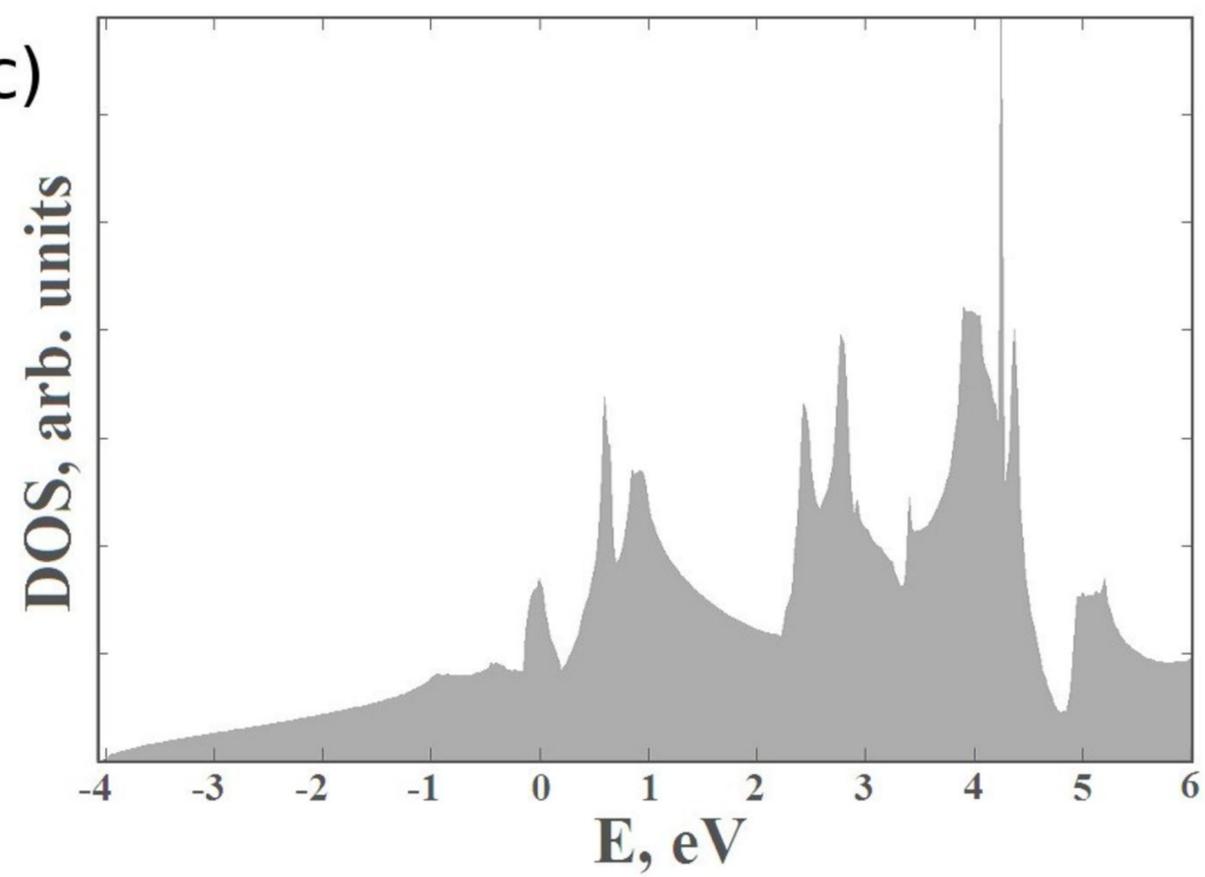 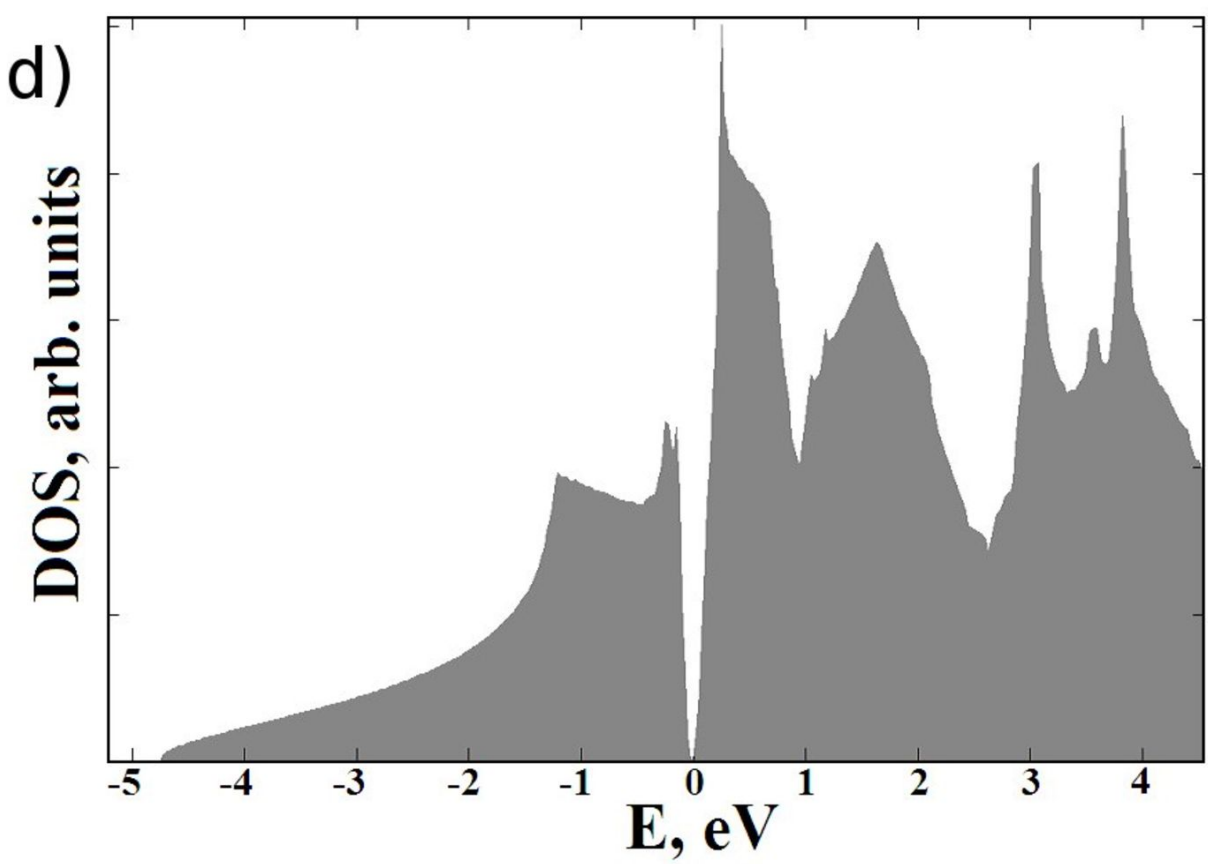

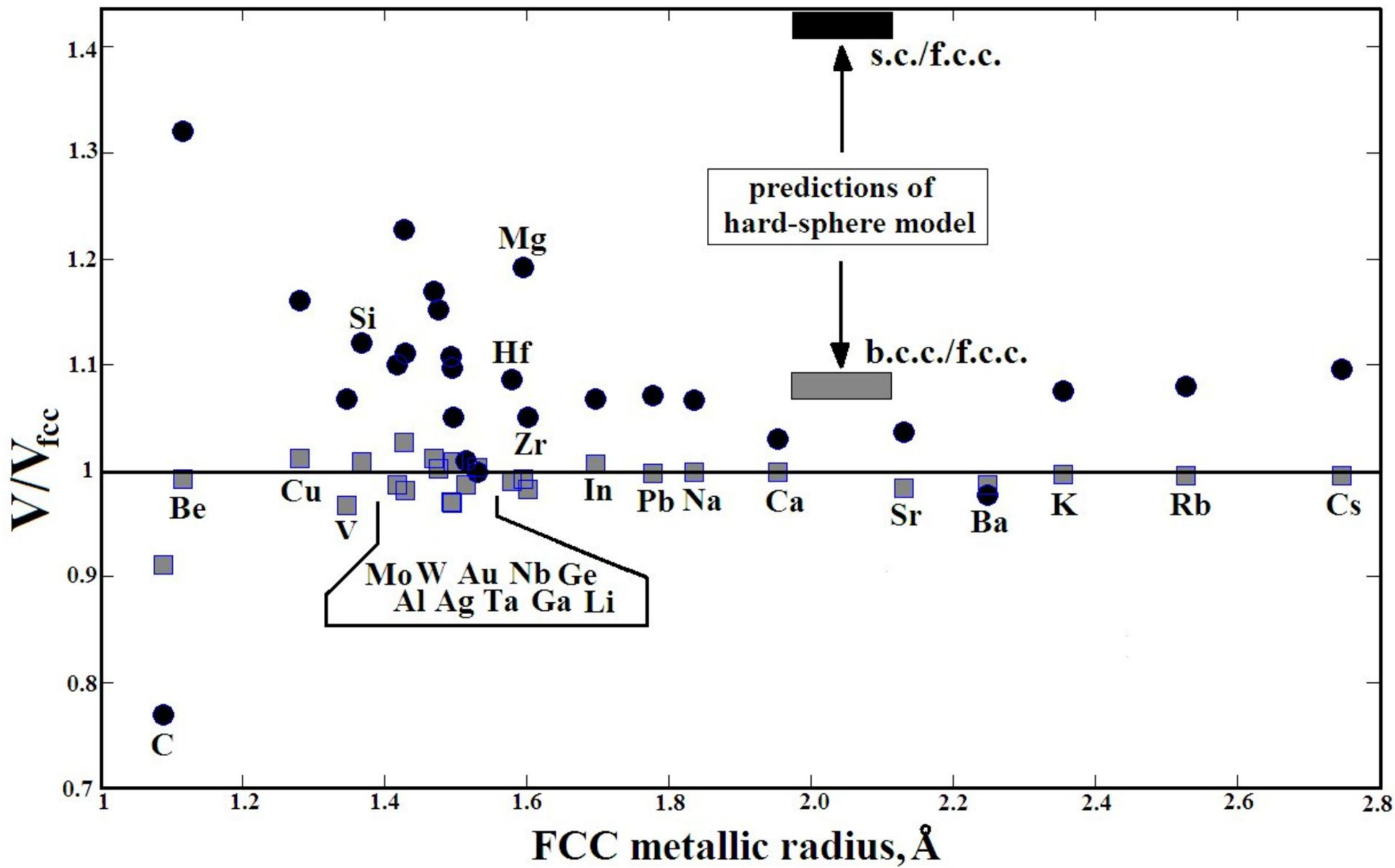

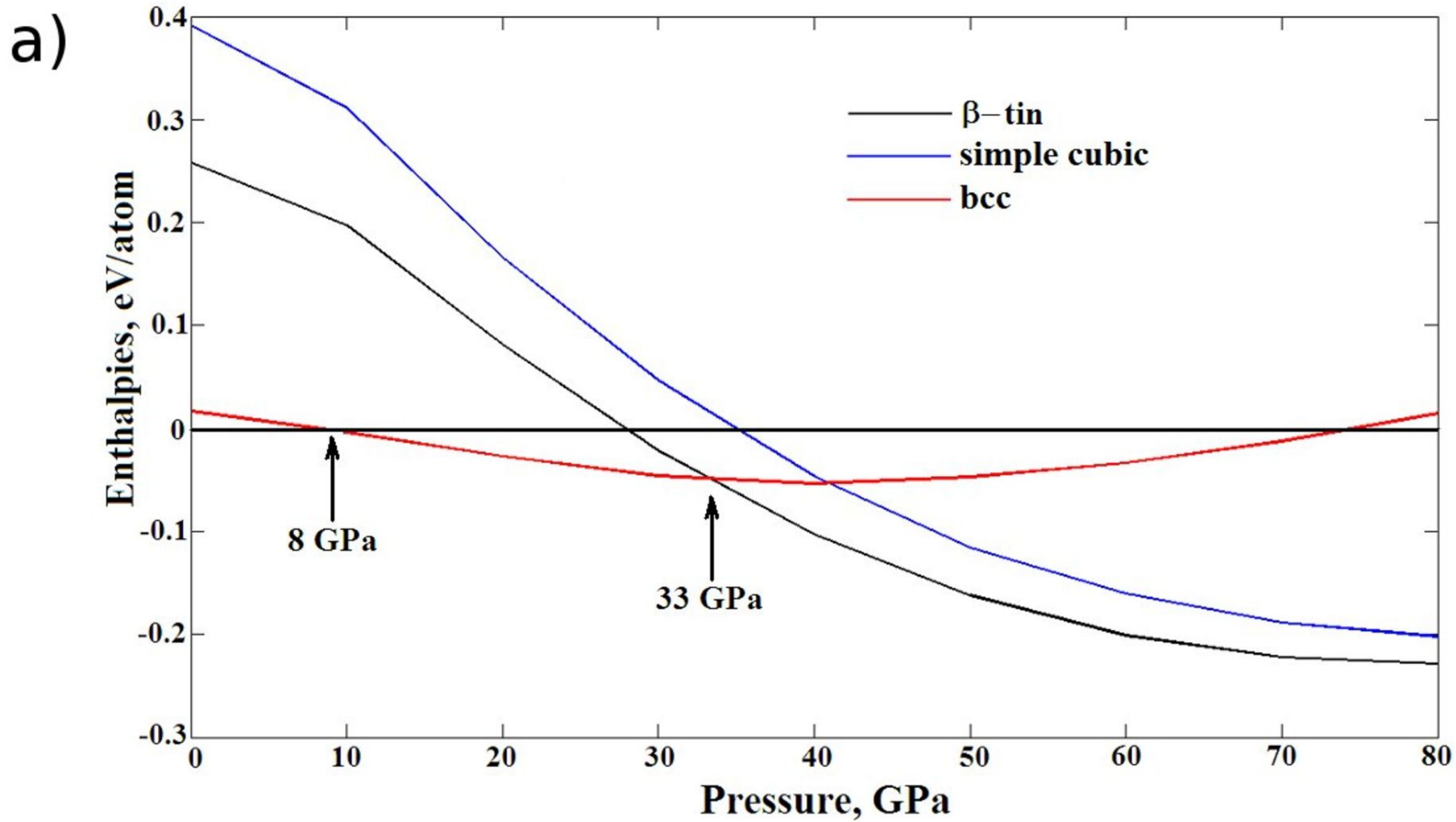

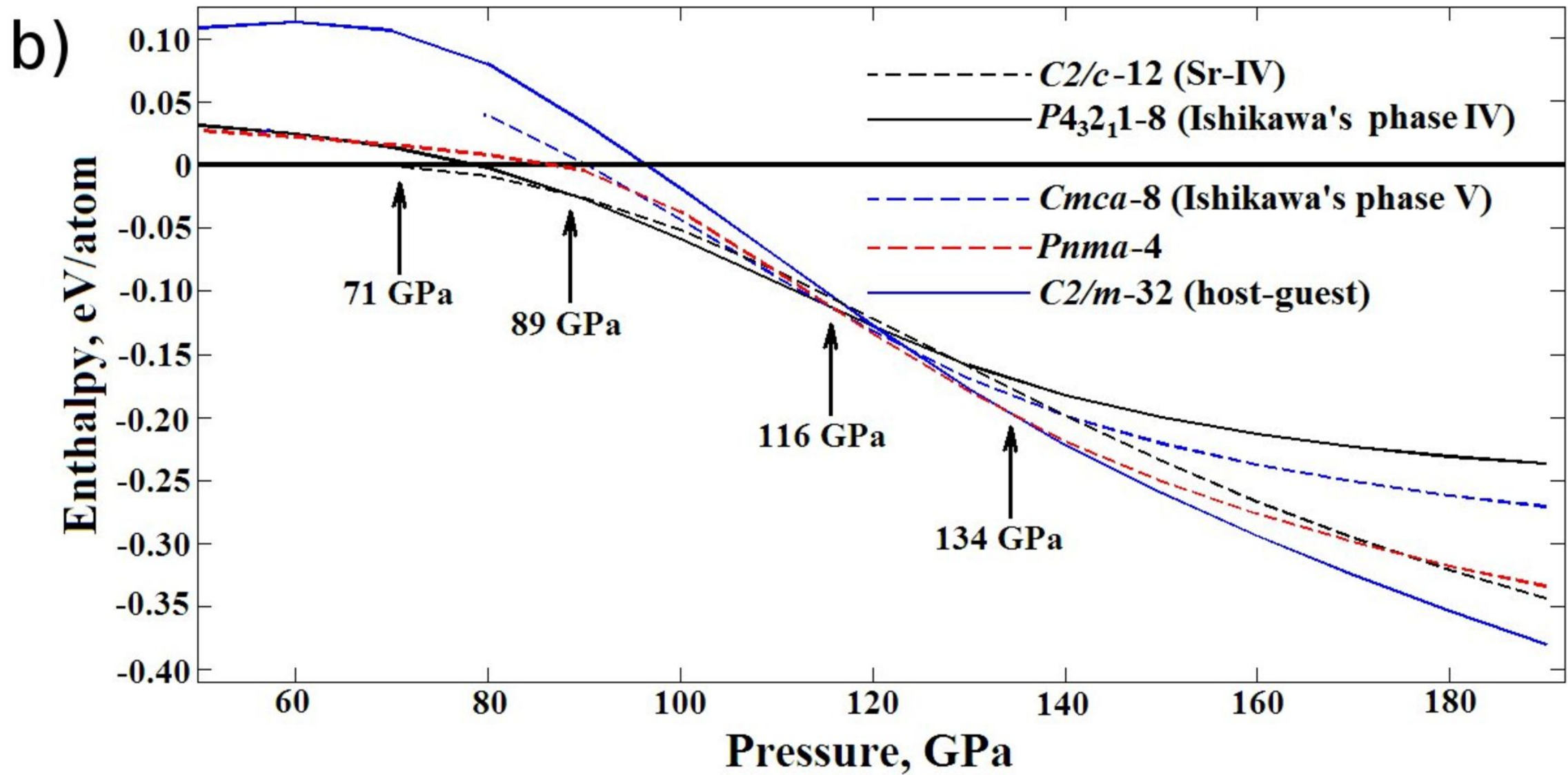

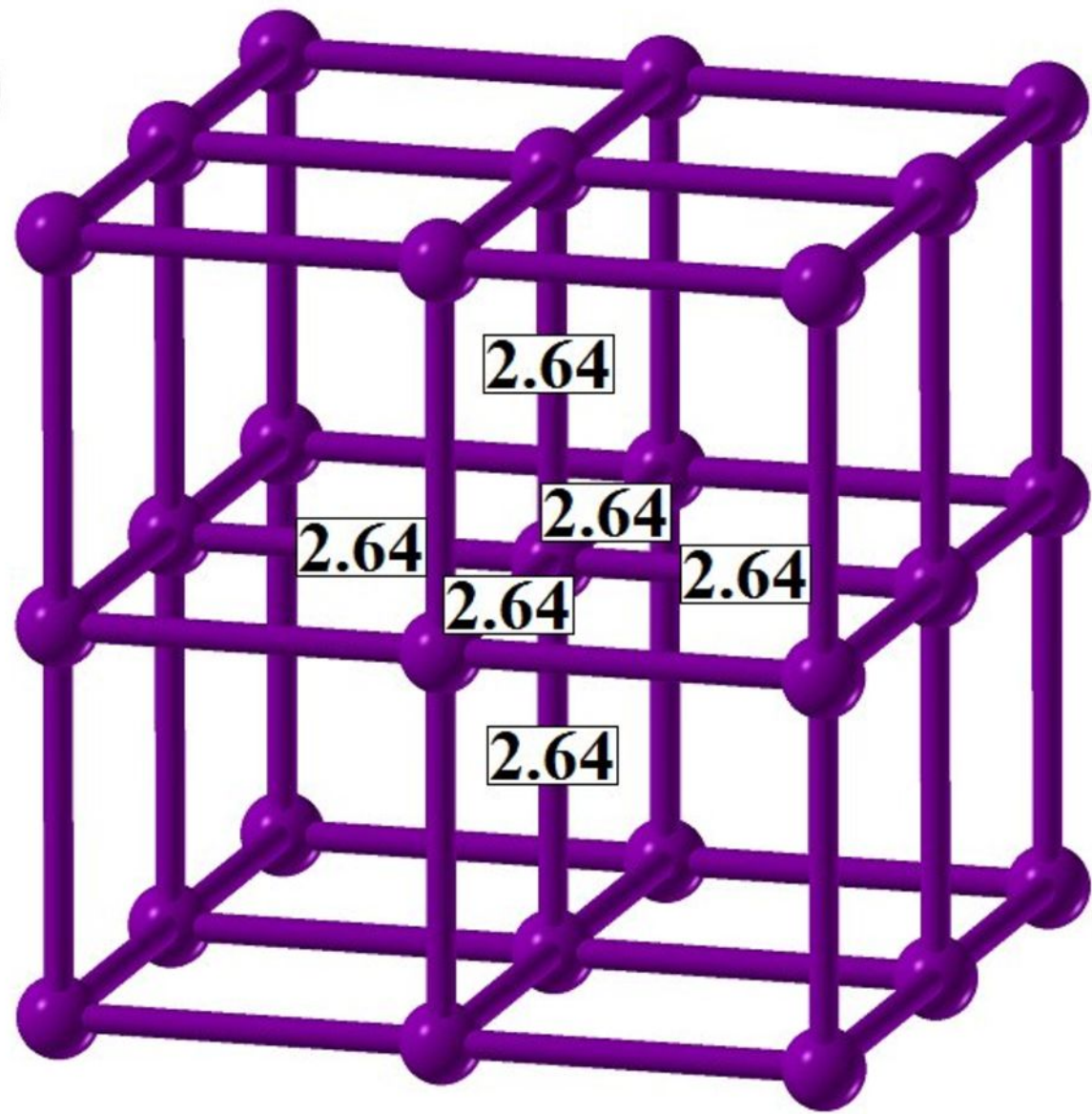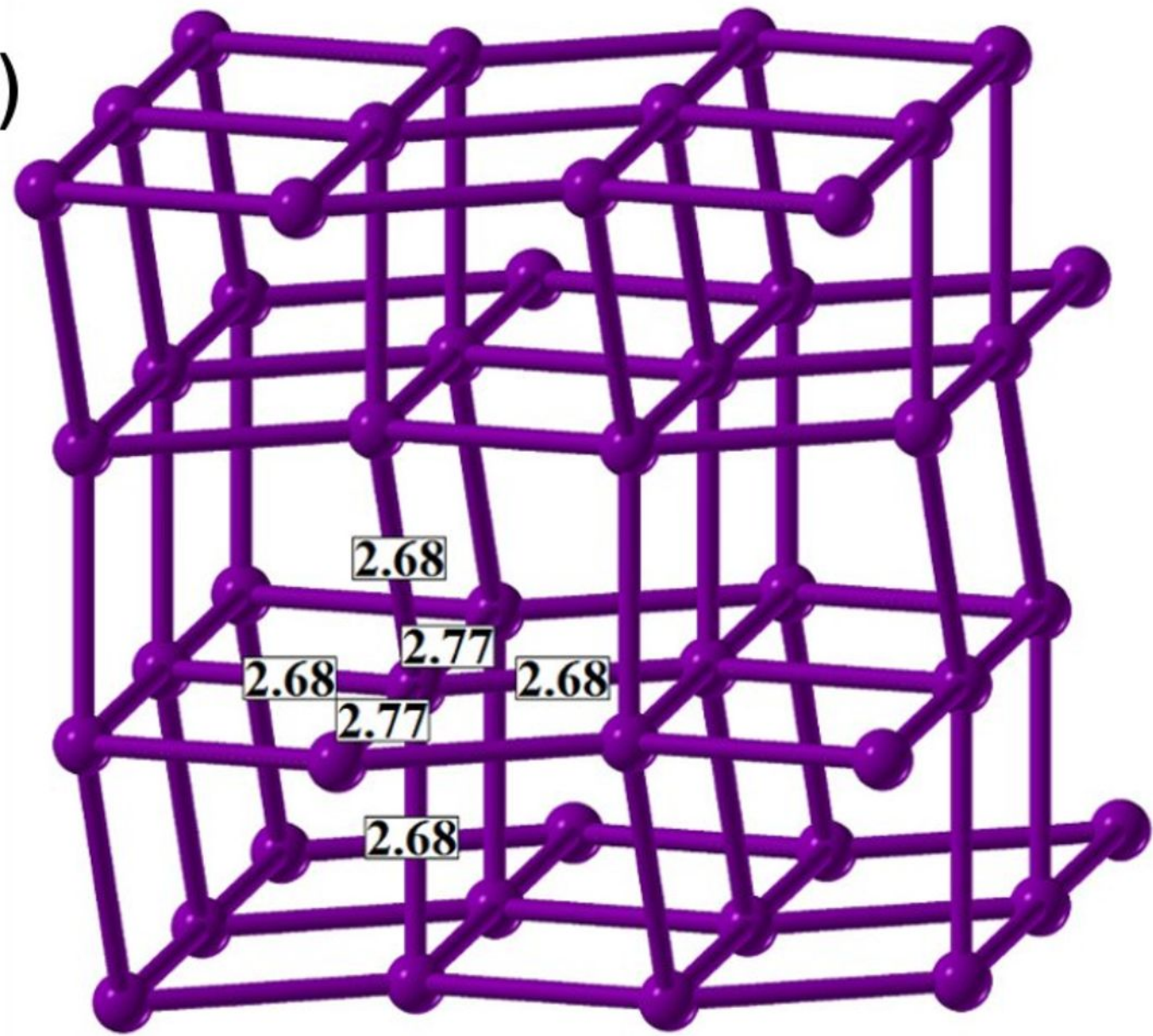

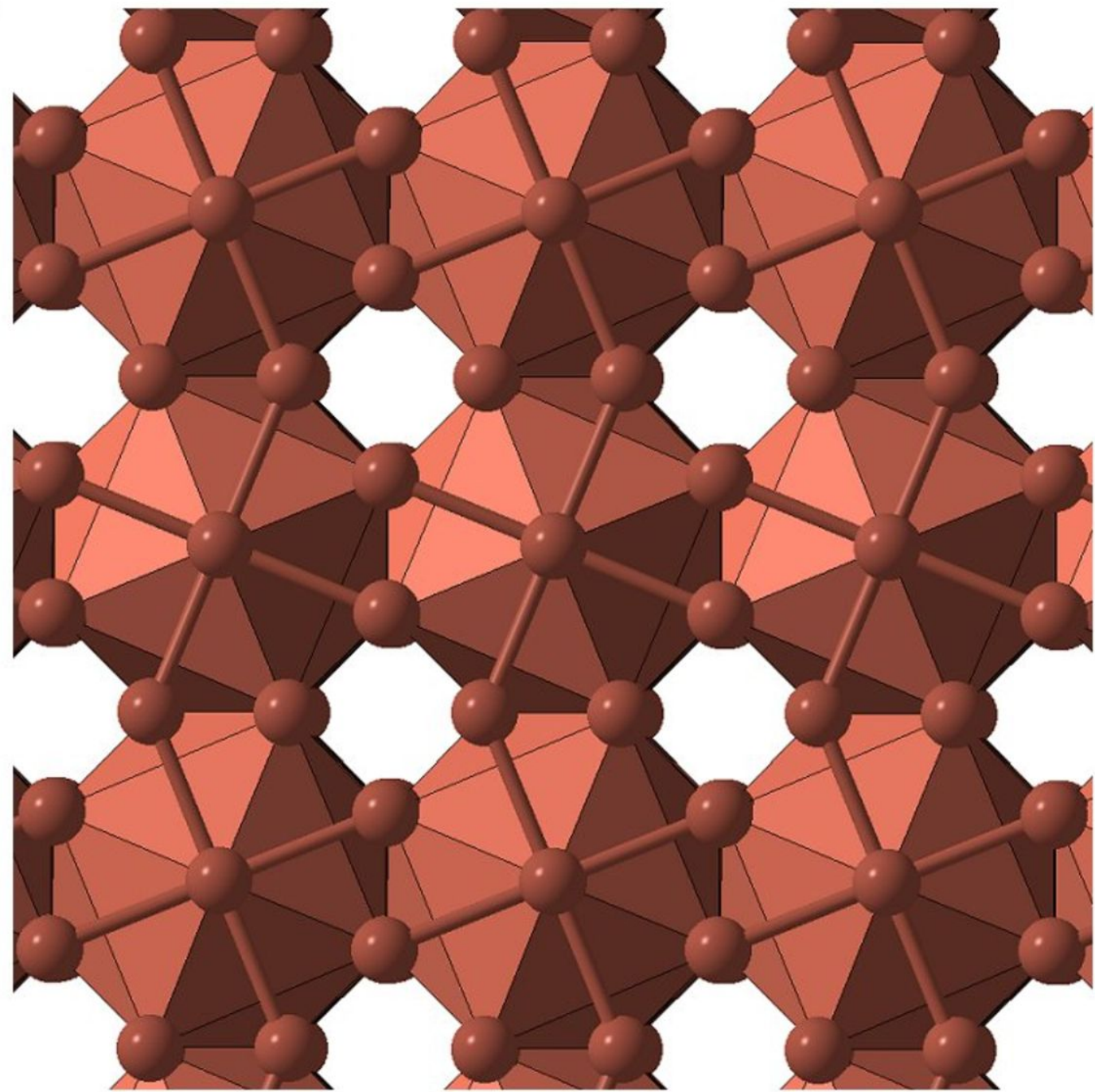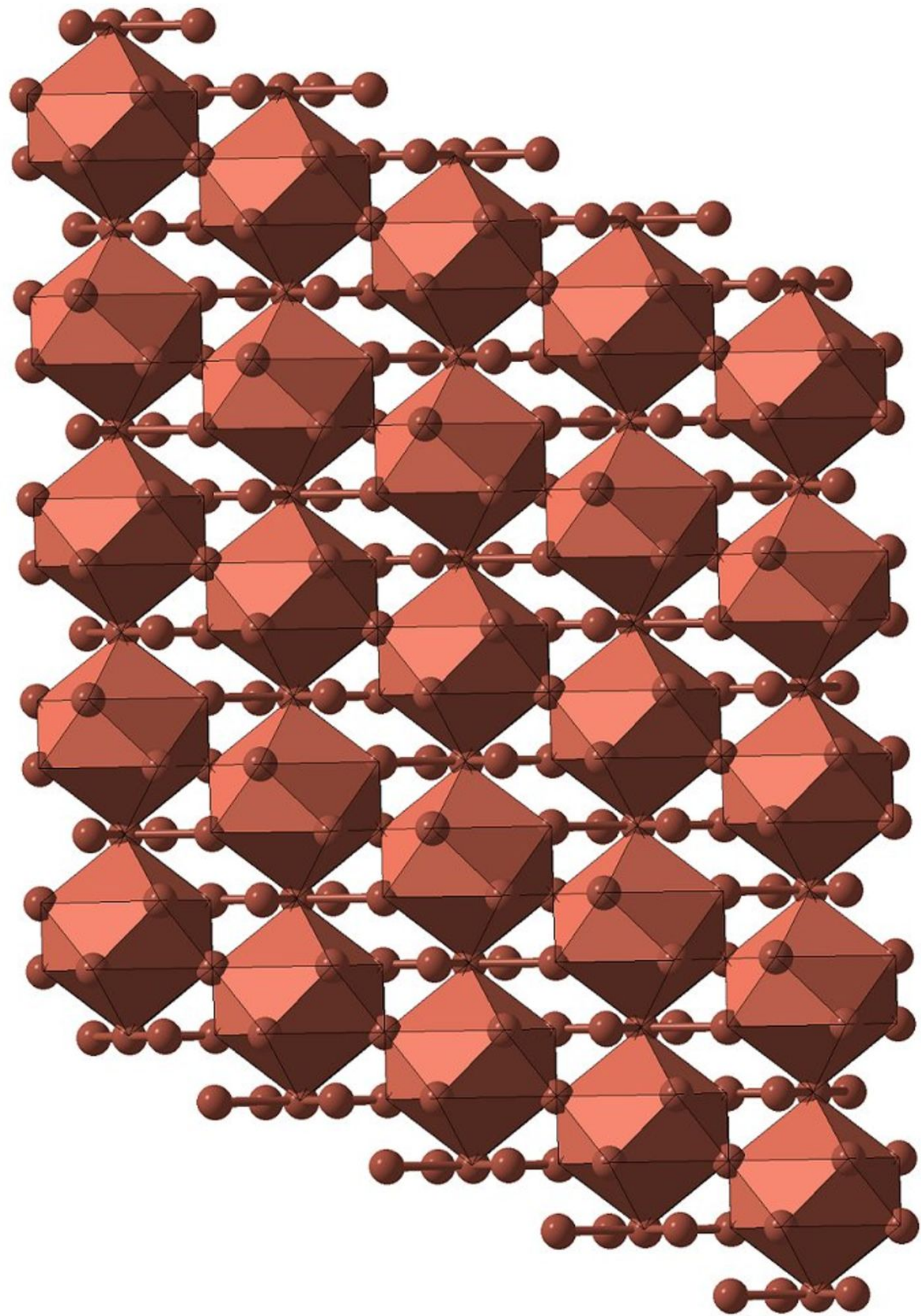

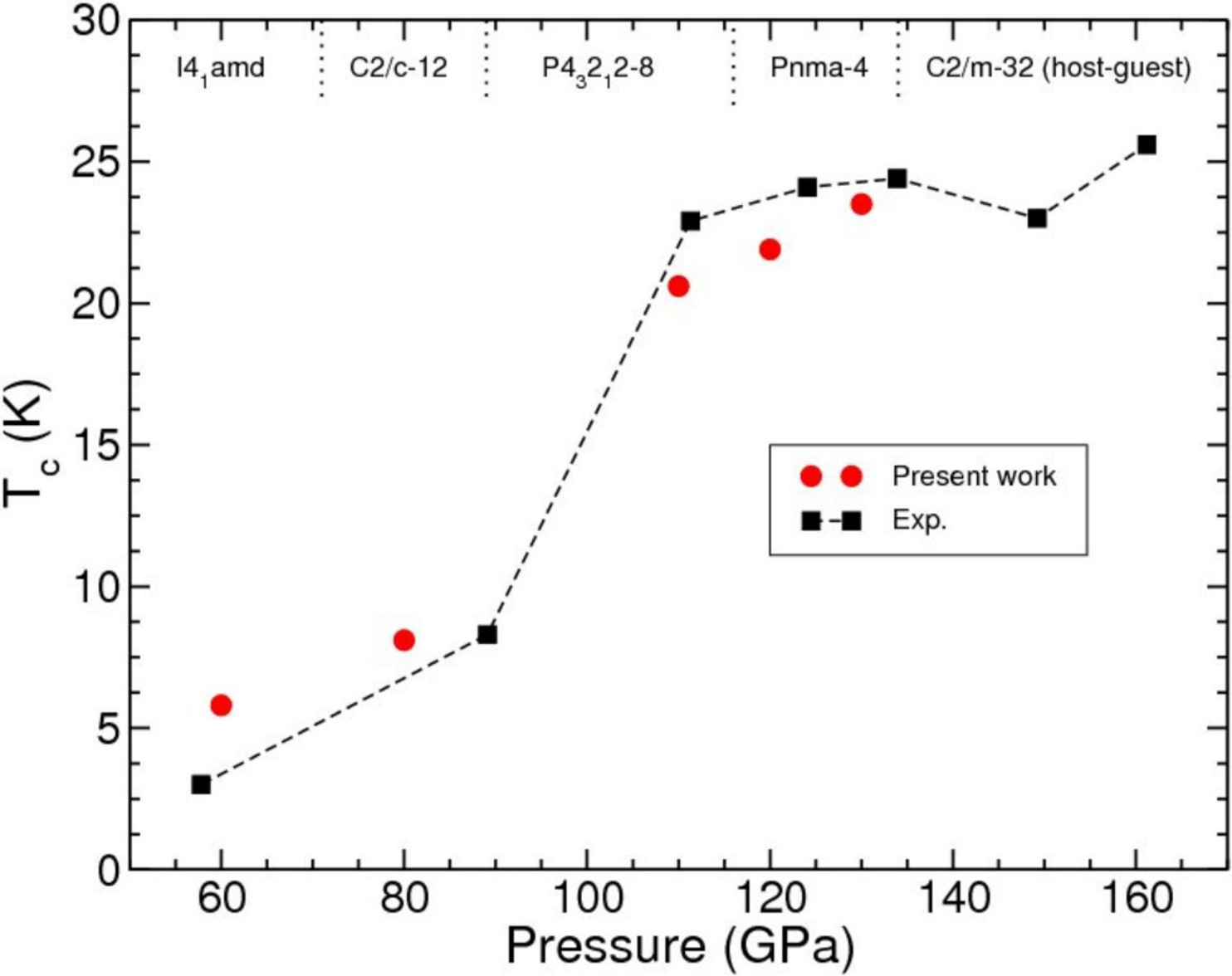

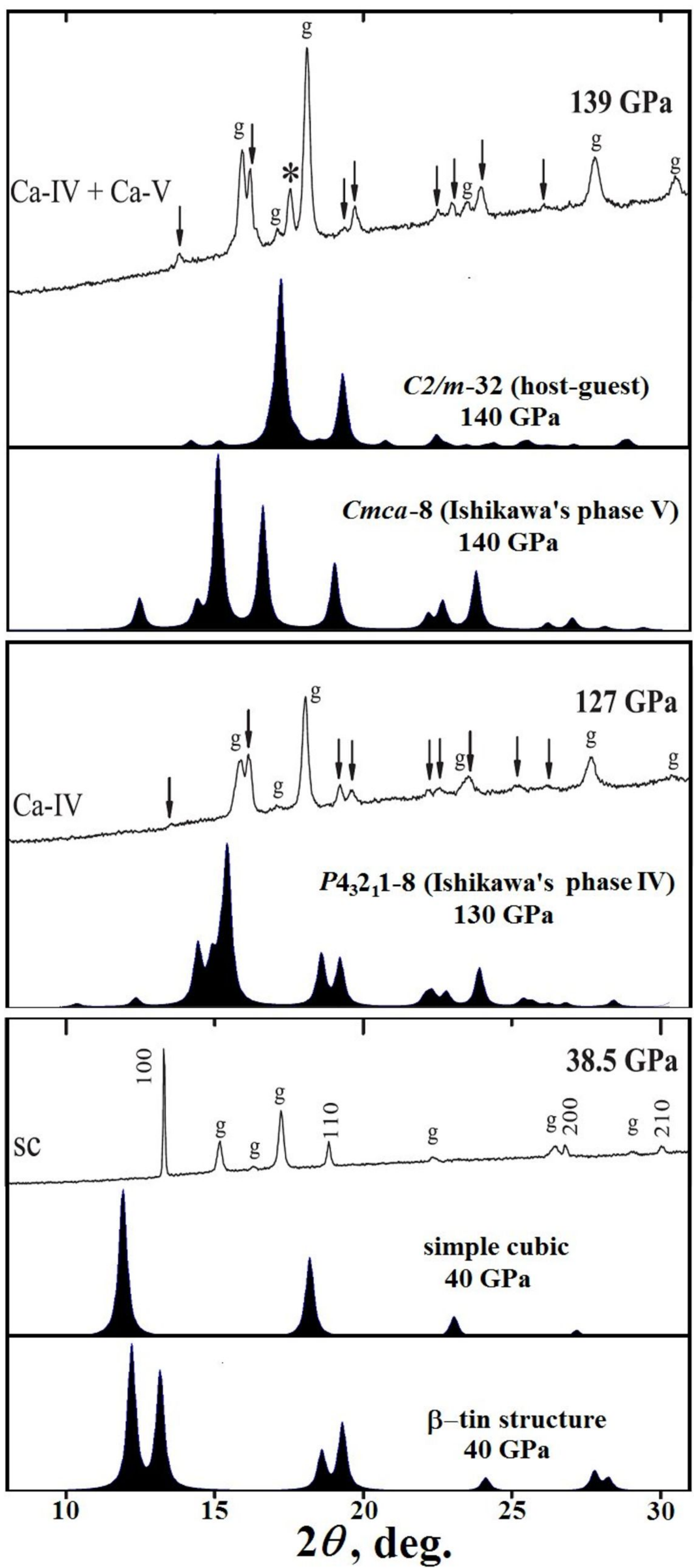

# Exotic behavior and crystal structures of calcium under pressure

**Artem R. Oganov[1,2], Ying Xu[3], Yanming Ma[3], Ion Errea[4,5], Aitor Bergara[4,5,6] and Andriy O. Lyakhov[1]**

[1]Department of Geosciences, Department of Physics and Astronomy, and New York Center for Computational Sciences, Stony Brook University, Stony Brook, NY 11794-2100, USA,
[2]Geology Department, Moscow State University, 119992 Moscow, Russia,
[3]National Lab of Superhard Materials, Jilin University, Changchun 130012, P.R. China,
[4]Materia Kondentsatuaren Fisika Saila, Zientzia eta Teknologia Fakultatea, Euskal Herriko Unibertsitatea, 644 Postakutxatila, 48080 Bilbo, Basque Country, Spain,
[5]Donostia International Physics Center (DIPC), Paseo de Manuel Lardizabal, 20018, Donostia, Basque Country, Spain,
[6]Centro Fisica de Materiale CSIC-UPV/EHU, 1072 Posta kutxatila, E-20080 Donostia, Basque Country, Spain.

Figs. 1-3 show the structures of the found phases of calcium under pressure, whose details are given in Table 1. Table 2 summarizes the pressure-induced transitions of Ca, Sr and Ba. Experimental results are compared with the present theoretical predictions. The similarity between the phase diagram of Sr and the predicted for Ca is clear.

Fig. 4 describes the density of states (DOS) of Ca at 50 GPa in both *sc* and β-tin phases. It is clear from the figure that a Peierls transition drives the distortion of the former, as it can be deduced from the energy-lowering opening of a (pseudo)gap at the Fermi level in the β-tin phase.

Fig. 5 displays the phonon spectrum of Ca in the β-tin phase at 60 GPa. The area of each circle is proportional to the partial contribution to the electron-phonon coupling

$$\lambda_{qv} \propto \gamma_{qv} / \omega^2_{qv} \qquad (1)$$

where $\gamma_{qv}$ is the linewidth of the mode $v$ at point **q** associated to the electron-phonon interaction and $\omega_{qv}$ its frequency. The Eliashberg function depicted in the right panel, together with the phonon density of states (PDOS), is calculated as

$$\alpha^2 F(\omega) = \frac{1}{2\pi N(0)} \sum_{qv} \frac{\gamma_{qv}}{\omega_{qv}} \delta(\omega - \omega_{qv}) \qquad (2)$$

The mode with largest contribution corresponds to a softenend optical mode. This is clear from the large circles of the figure and the lack of correspondence between the Eliashberg function and the PDOS. This mode is the transverse, unstable and very anharmonic mode at M found in *sc*-Ca. Such correspondence can be deduced from the eigenvectors of the dynamical matrices. Table 3 summarizes the results obtained for the electron-phonon parameters that enter intoMcMillan equation for different pressures and phases. The logarithmic average frequency is calculated as

$$\omega_{\log} = \exp\left(\frac{2}{\lambda} \int_0^\infty d\omega \frac{\alpha^2 F(\omega)}{\omega} \ln \omega \right) \qquad (3)$$

where

$$\lambda = 2\int_0^\infty d\omega \frac{\alpha^2 F(\omega)}{\omega} \qquad (4)$$

is the electron-phonon coupling constant.

SUPPORTING FIGURE LEGENDS:

**Fig. 1.** $C2/c$-12 (Sr-IV) structure in two views (a,b and c,d, respectively) at 50 GPa and 100 GPa (from left to right). At 50 GPa the structure collapses into β-tin structure from which it appears as a result of a symmetry lowering second-order phase transition at 71 GPa. This figure shows a close relationship between these two structures.

**Fig. 2.** $P4_32_12$-8 (a,b) and $Cmca$-8 (c,d) structures of Ishikawa. These structures are very similar and can be considered as heavy distortions of the s.c. structure, which is especially clearly seen in (b) and (c). The structure shown in (c,d) is metastable (i.e. not a ground state) at $T$=0 K.

**Fig. 3.** $Pnma$-4 structure at 150 GPa. The structure can be described as consisting of layers (two layers are shown here). This description is only formal, as distances within the layer (2.29 Å) are only slightly shorter than between the layers (2.31 and 2.32 Å).

**Fig. 4.** Comparison between the DOS of $sc$ and $I4_1/amd$ phases at 50 GPa. In the β-tin phase a pseudogap develops close to the Fermi level, which is a characteristic of the Peierls transition.

**Fig. 5.** Phonon spectrum of calcium at 60 GPa in the $I4_1/amd$ structure (left panel). The area of each circle depicted is proportional to the partial electron-phonon coupling $\lambda_{q\nu}$. The PDOS (solid line) and the Eliashberg function $\alpha^2 F(\omega)$ (dashed line) in arbitrary units are plotted together with the integrated $\lambda(\omega) = 2\int_0^\omega d\omega' \alpha^2 F(\omega')/\omega'$ (dash-dotted line).

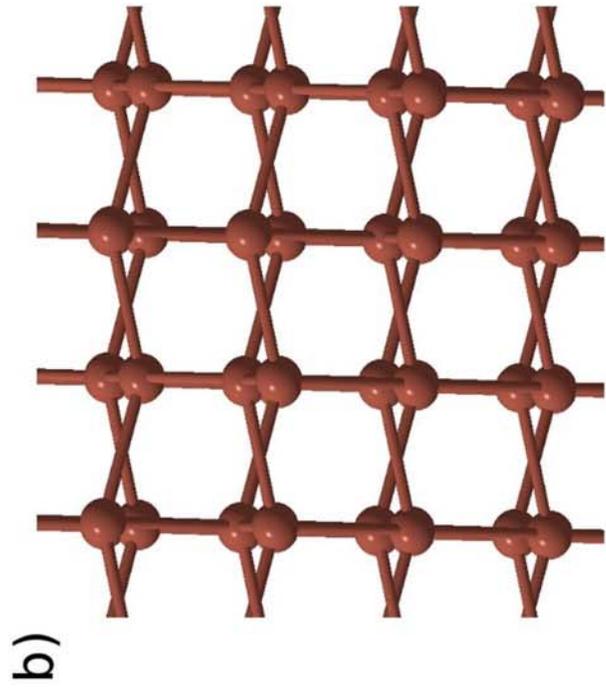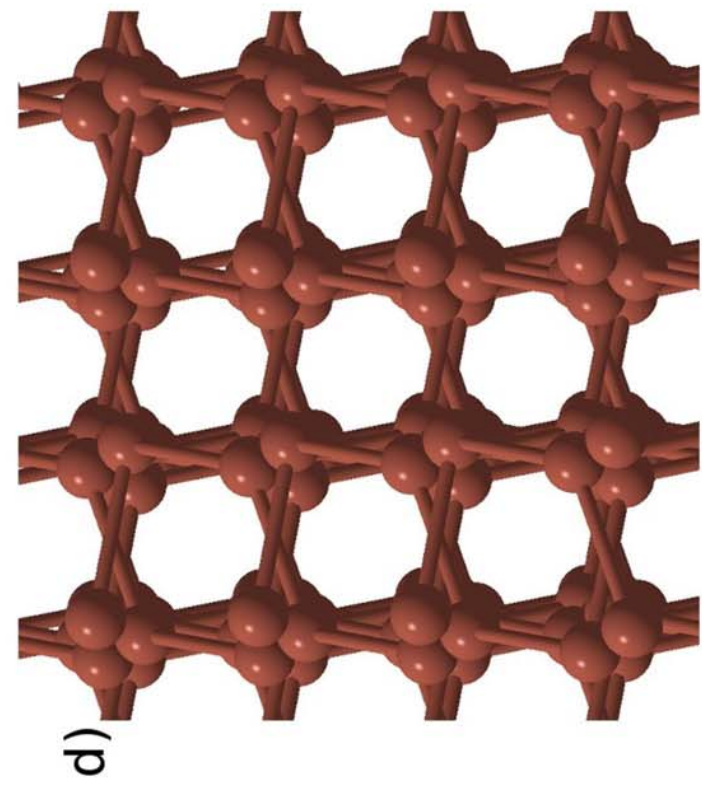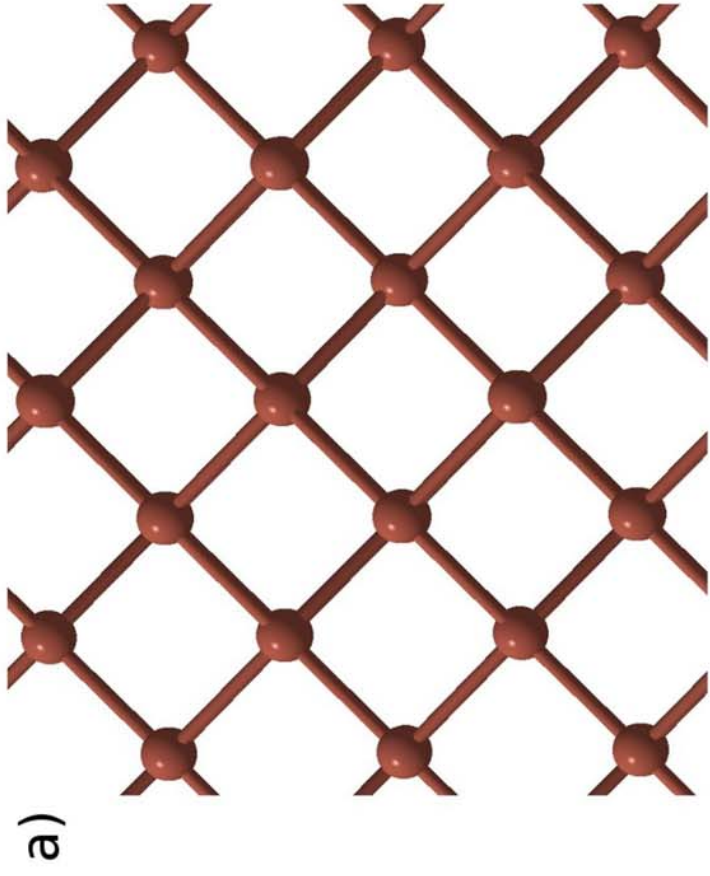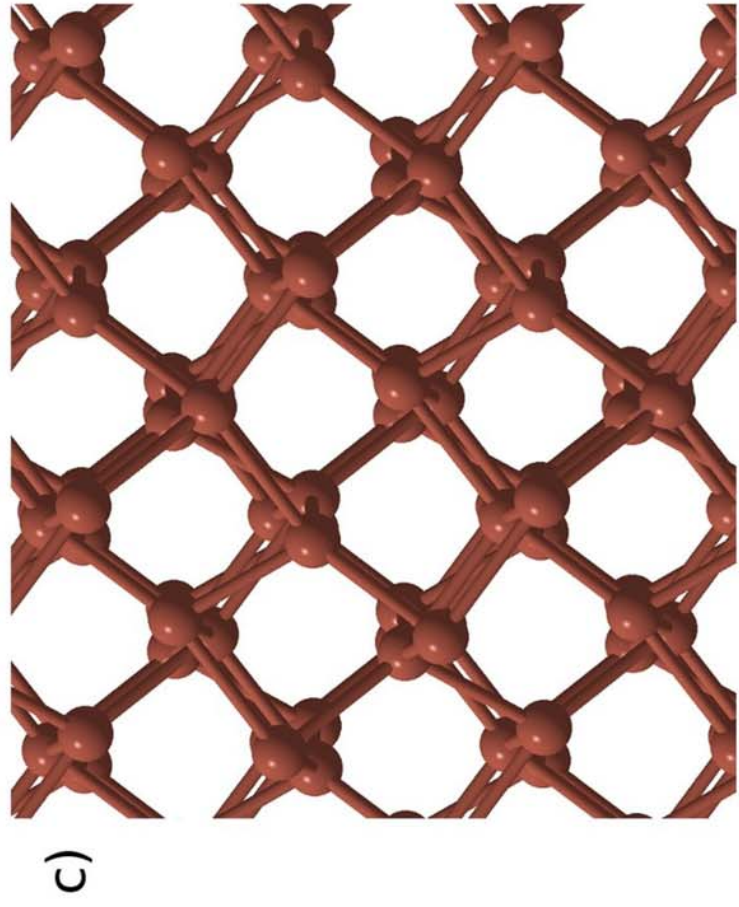

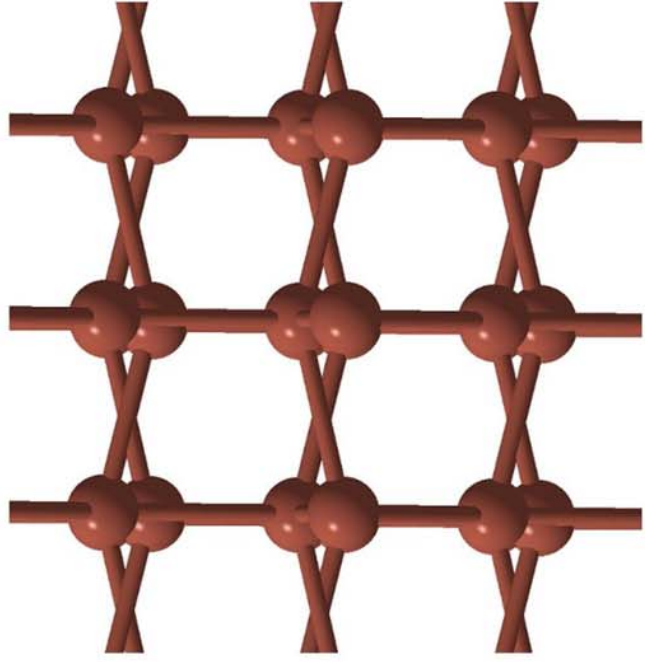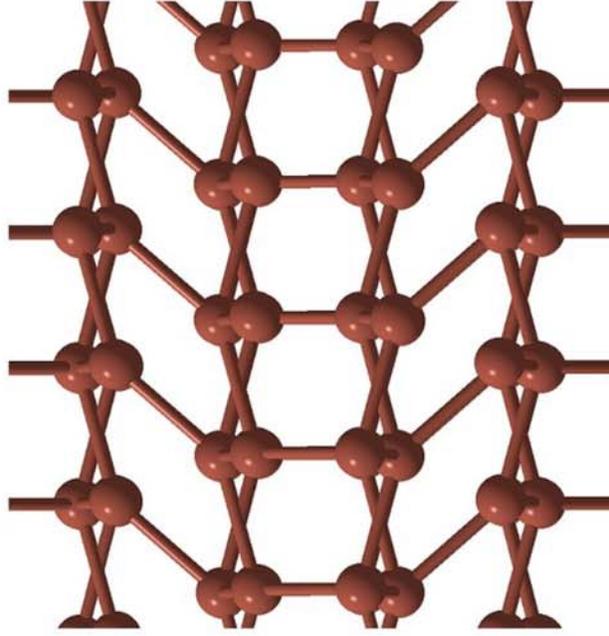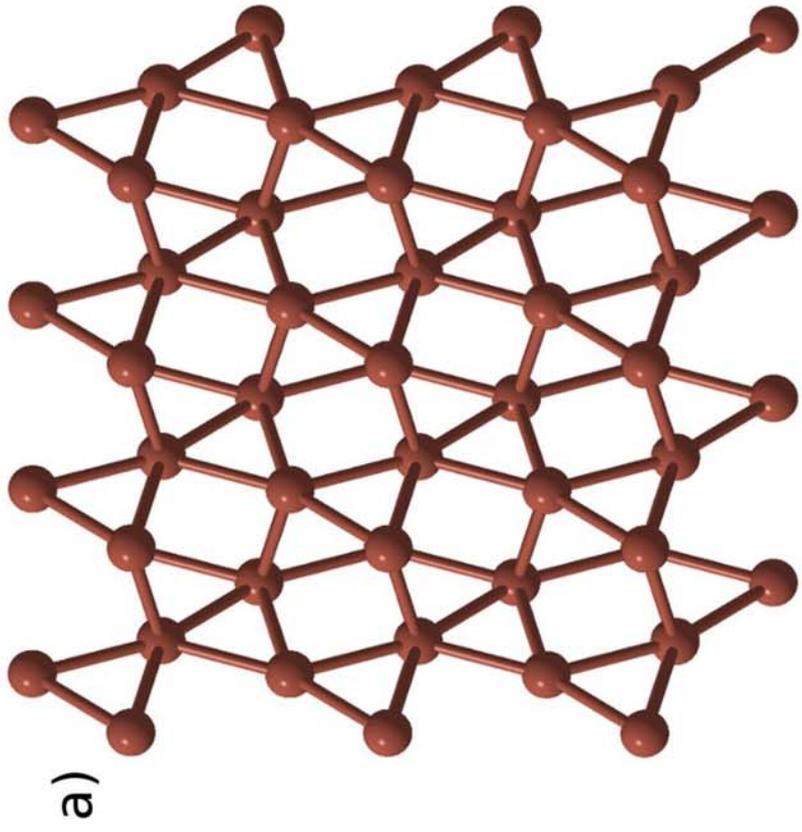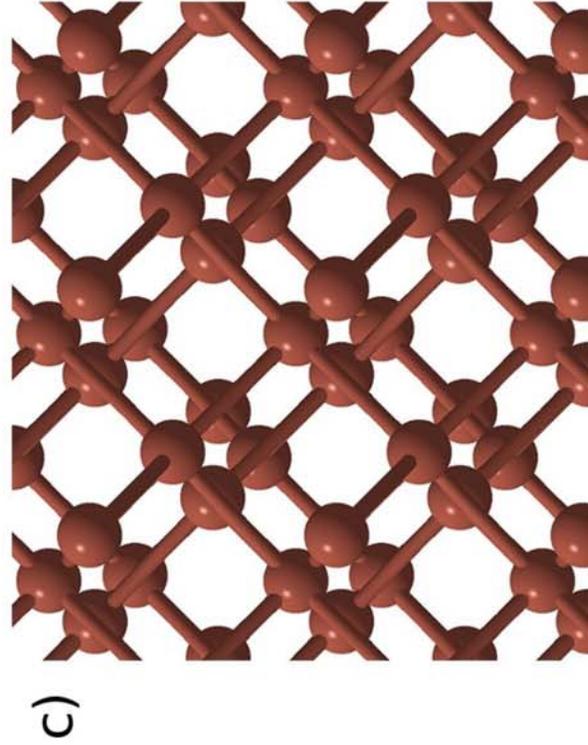

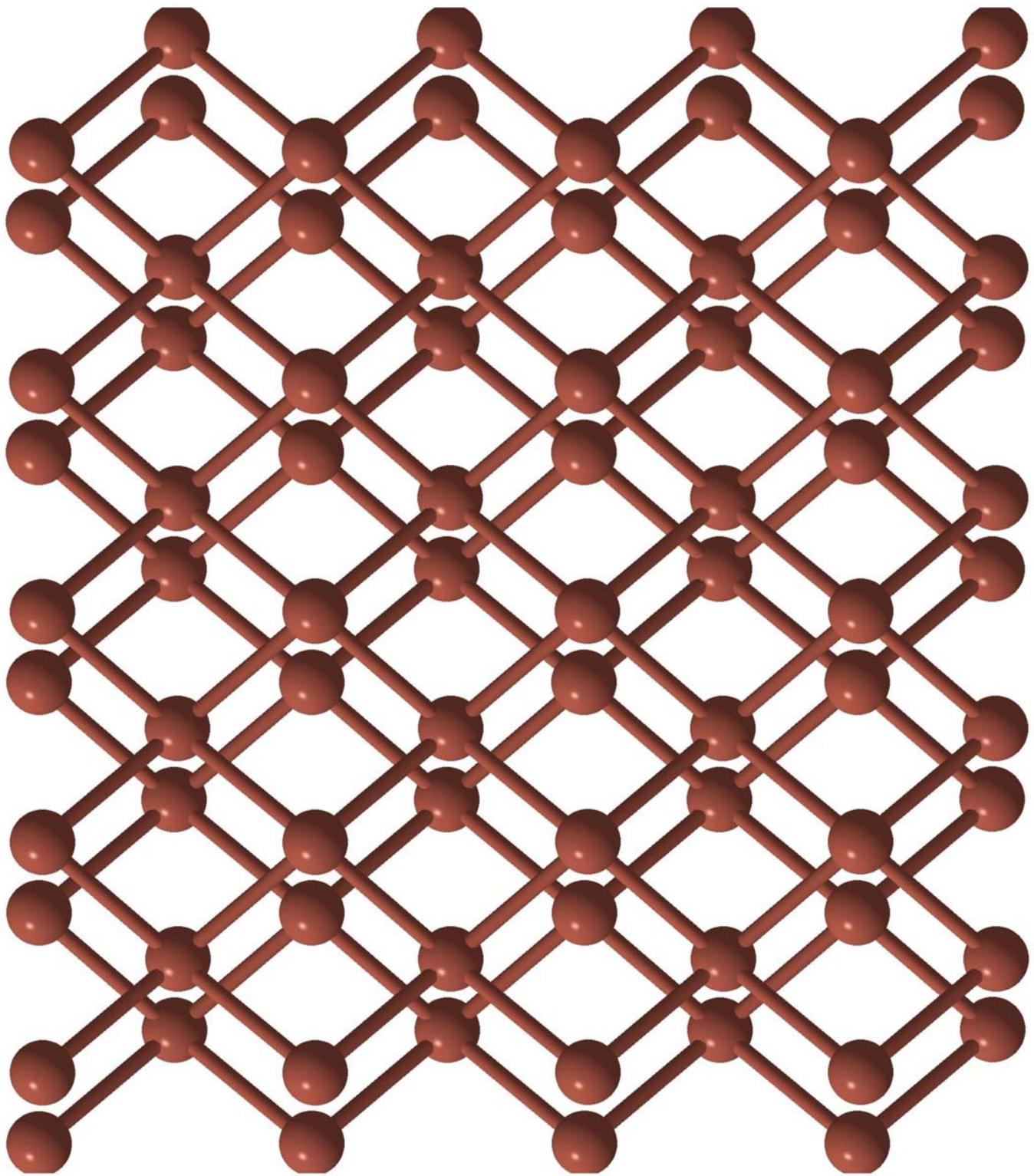

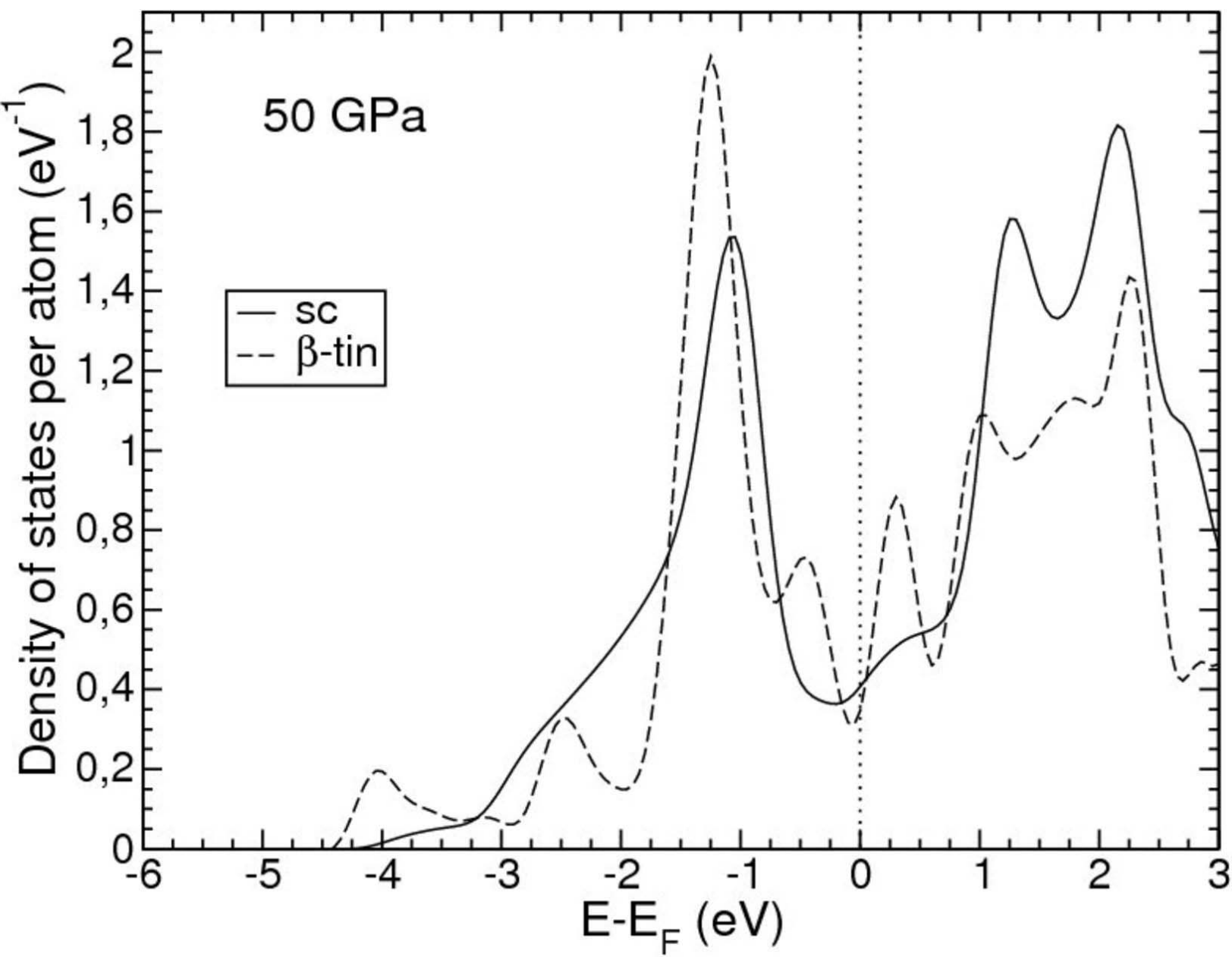

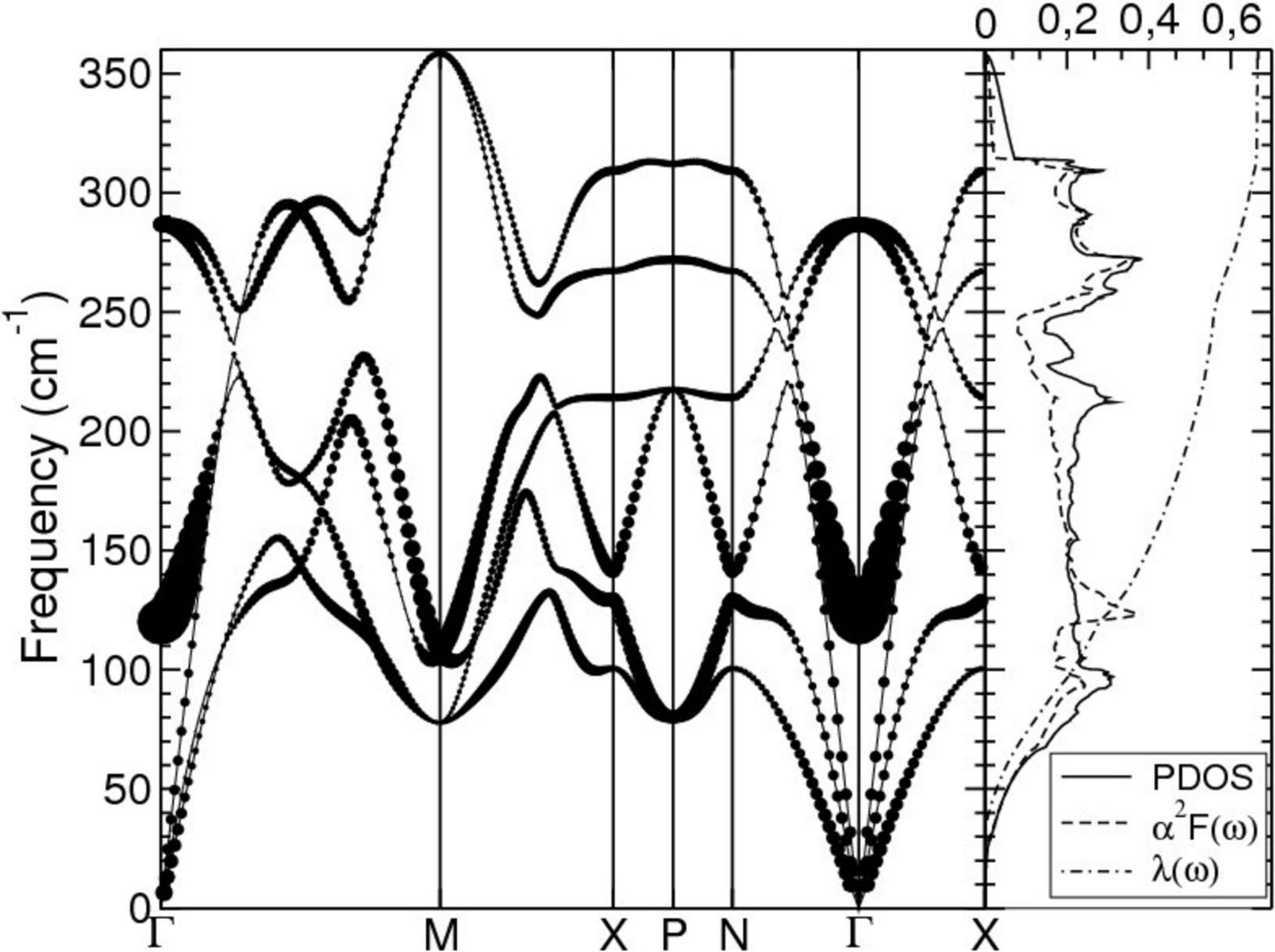

**Table 1.** Structural parameters at 130 GPa of some of the structures studied. The unit cell parameters are given in Å. The Wyckoff orbits are indicated together with the position in the cell of one of its representatives.

| | Wyckoff position | x | y | z |
|---|---|---|---|---|
| β-tin phase ($I4_1/amd$-4), space group origin 2 | | | | |
| a = b = 4.3447, c = 2.4286 | | | | |
| Ca | 4b | 0.00000 | 0.25000 | 0.37500 |
| Sr-IV type phase ($C2/c$-12) | | | | |
| a = 6.2387, b = 6.2205, c = 4.4055 (Å), β = 130.35° | | | | |
| Ca1 | 4e | 0.00000 | 0.80441 | 0.25000 |
| Ca2 | 8f | 0.78656 | 0.59057 | 0.42230 |
| $P4_32_12$-8 (Isikawa's phase IV) | | | | |
| a = b = 3.1219, c = 9.0619 (Å) | | | | |
| Ca | 8b | 0.82148 | 0.48884 | 0.09360 |
| $Pnma$-4 | | | | |
| a = 4.3924, b = 3.3952, c = 2.8968 (Å) | | | | |
| Ca | 4c | 0.67192 | 0.75000 | 0.60785 |
| $C2/m$-32 (host-guest) | | | | |
| a = 10.0464, b = 8.2557, c = 4.4698 (Å), β = 112.37° | | | | |
| Ca1 | 4e | 0.25000 | 0.25000 | 0.00000 |
| Ca2 | 4g | 0.00000 | 0.24936 | 0.00000 |
| Ca3 | 4h | 0.00000 | 0.35380 | 0.50000 |
| Ca4 | 4i | 0.54921 | 0.50000 | 0.79035 |
| Ca5 | 8j | 0.66766 | 0.64694 | 0.49543 |
| Ca6 | 4i | 0.78372 | 0.50000 | 0.20683 |
| Ca7 | 4i | 0.88468 | 0.50000 | 0.79557 |
| $I4/mcm$-32 (host-guest) | | | | |
| a = b = 5.8377, c = 10.0493 (Å) | | | | |
| Ca1 | 16l | 0.14756 | 0.35244 | 0.16727 |
| Ca2 | 8h | 0.85524 | 0.35524 | 0.00000 |
| Ca3 | 4a | 0.50000 | 0.50000 | 0.75000 |
| Ca4 | 4c | 0.50000 | 0.50000 | 0.00000 |
| $C2/c$-32 (host-guest) | | | | |
| a = 8.9121, b = 7.9852, c = 11.5509 (Å), β = 107.39° | | | | |
| Ca1 | 8f | 0.69106 | 0.40326 | 0.78606 |
| Ca2 | 8f | 0.60496 | 0.75384 | 0.87754 |
| Ca3 | 8f | 0.52937 | 0.11675 | 0.61746 |
| Ca4 | 8f | 0.83392 | 0.92019 | 0.44795 |
| $C2/c$-24 | | | | |
| a = 7.6867, b = 4.4218, c = 8.3884 (Å), β = 114.23° | | | | |
| Ca1 | 8f | 0.77472 | 0.24566 | 0.80848 |
| Ca2 | 8f | 0.02764 | 0.82753 | 0.90171 |
| Ca3 | 8f | 0.89154 | 0.33840 | 0.59753 |
| $Cmca$-8 (Isikawa's phase V) | | | | |
| a = 4.36926, b = 4.56578, c = 4.38753 (Å) | | | | |
| Ca | 8f | 0.00000 | 0.65701 | 0.30797 |
| $Cmca$-16 | | | | |
| a = 4.3221, b = 7.9473, c = 5.2744 (Å) | | | | |
| Ca1 | 8f | 0.00000 | 0.7880 | 0.0533 |
| Ca2 | 8f | 0.00000 | 0.0235 | 0.2940 |

**Table 2. Summary of high-pressure behaviour of Ca, Sr and Ba. For Ca the theoretical results presented here are given in parenthesis.**

| | *fcc* | *bcc* | *sc* derivative structures: | | | | Host-guest *C2/m*-32 | *hcp* |
|---|---|---|---|---|---|---|---|---|
| | | | $I4_1/amd$, (β-tin) | *C2/c*-12 (Sr-IV) | $P4_32_12$-8 | *Pnma*-4 | | |
| Ca | 0-19.5 GPa | 19.5-32 GPa | 32-113 GPa (*sc*) | | 113-139 GPa | ? | > 139 GPa | ? |
| | (0-8 GPa) | (8-33 GPa) | (33-71 GPa) | (71-89 GPa) | (89-116 GPa) | (116-134 GPa) | (134-564 GPa) | (>564 GPa) |
| Sr | 0-3.5 GPa | 3.5-26 GPa | 26-35 GPa | 35- 26.3 GPa | | | > 46.3 GPa | ? |
| Ba | | 0-5.5 GPa∗ | | | | | 12.6-45 GPa | > 45 GPa |

* at 5.5-12.6 GPa Ba adopts the hcp structure (re-entrant above 45 GPa)

Table 3. Summary of the calculated magnitudes entering McMillan equation in the different phases.

| | Pressure (GPa) | $\omega_{\log}$ (K) | λ | Tc(K) |
|---|---|---|---|---|
| $I4_1/amd$-4 | 60 | 188.7 | 0.67 | 5.8 |
| $C2/c$-12 (Sr-IV) | 80 | 80.4 | 1.33 | 8.1 |
| $P4_32_12$-8 | 110 | 140.7 | 2.06 | 20.6 |
| Pnma-4 | 120 | 333.0 | 0.96 | 21.9 |
| Pnma-4 | 130 | 348.5 | 0.98 | 23.5 |